\documentclass{llncs}

\usepackage{amsmath}
\usepackage{latexsym}

\input{psfig.sty}

\newcommand\alld{\texttt{all\-different}}

\begin{document}

\bibliographystyle{plain}

\pagestyle{empty}

\mainmatter

\title{The \alld\ Constraint: A Survey}


\author{W.J. van Hoeve\inst{}}


\institute{CWI, P.O. Box 94079, 1090 GB Amsterdam, The Netherlands\\
\email{wjvh@cwi.nl}\\
\texttt{http://www.cwi.nl/\~{ }wjvh}}

\maketitle

\begin{abstract}
The constraint of difference is known to the constraint programming community 
since Lauriere introduced \textsc{Alice} \cite{lauriere} in 1978. Since then,
several solving strategies have been designed for this constraint. In this 
paper we give both a practical overview and an abstract comparison of these
different strategies.
\end{abstract}

\section{Introduction}
Many problems from combinatorial optimization can be modeled and solved using
techniques from Constraint Programming \cite{MS:prog98,vanh:chip}. One of the
constraints that arises naturally in these models is the \alld\ constraint,
which states that all variables in this constraint must be pairwise different.
In Example~\ref{ex:speech}, a scheduling problem is modeled using the \alld\
constraint.

\begin{example}[Scheduling of speeches]\label{ex:speech}
Consider the following simple scheduling problem, adapted from Puget 
\cite{puget98},
where a set of speeches must be scheduled during one day. Each speech lasts 
exactly one hour (including questions and a coffee break), and only one 
conference room is available. Furthermore, each speaker has other commitments,
and is available only for a limited fraction of the day. A particular instance
of this problem is given in Table~\ref{tb:times}, where the fractions are 
defined by an earliest and latest possible time slot. 
\begin{table}
\begin{center}
\caption{Time slots for the speakers}
\label{tb:times}
\begin{tabular}{|l|c|c|} \hline \hline 
Speaker & Earliest & Latest\\ \hline
Sebastian    & 3 & 6 \\ 
Fr\'ed\'eric & 3 & 4 \\ 
Jan-Georg    & 2 & 5 \\ 
Krzysztof    & 2 & 4 \\ 
Maarten      & 3 & 4 \\ 
Luca         & 1 & 6 \\ \hline \hline
\end{tabular}
\end{center}
\end{table}
This problem can be modeled as follows. We create one variable per speaker, 
whose value will be the period of his speech. The initial domains of the
variables will be the available time intervals as stated in Table 
\ref{tb:times}. Since two speeches cannot be held at the same time in the 
same conference room, the period for two different speakers must be different.
The constraints for this scheduling problem thus become:
\begin{displaymath}
\begin{array}{c}
x_1 \in [3,6], x_2 \in [3,4], x_3 \in [2,5], \\
x_4 \in [2,4], x_5 \in [3,4], x_6 \in [1,6], \\
$ {\tt alldifferent}$(x_1, x_2, x_3, x_4, x_5, x_6).
\end{array}
\end{displaymath}
\end{example}

To find a solution to a model as in the previous example, a constraint 
solver essentially builds a search tree from all possible 
variable values. In general, finding a solution for such problems is
${\cal NP}$-complete, and this search tree can grow extremely large. 
Therefore, strategies have been developed to prune parts of the search tree.
In Constraint Programming, these strategies mainly consist of the 
simplification
of the problem during the search for a solution. The techniques that are most
widely applied are so-called consistency techniques that can reduce the domains
of the variables, based on the constraints between them. Therefore, algorithms
that achieve some state of consistency are also called filtering algorithms.

This paper deals with consistency techniques or filtering algorithms that can 
be deduced from the \alld\ constraint. It turns out that there exist different
degrees of consistency, each degree allowing more or less values in the 
variable domain. In general it takes more time to obtain a stronger 
consistency than to obtain a weaker consistency. So with more effort, one could
remove more values. Therefore, for each individual problem one has to make a 
trade-off between the effort (time) and the gain (domain shrinking) when 
choosing a particular consistency to achieve.

\subsection{Overview}
The different degrees of consistency will be defined in Section~\ref{s:pre}, 
together with some more preliminaries. Then each of the Sections~\ref{s:dis}
up to~\ref{s:arc} will treat one consistency technique. These sections are 
ordered in increasing strongness of the considered consistency. The treatment 
consists of a description of the particular consistency with respect to the 
\alld\ constraint, together with an algorithm that achieves this consistency.
Finally, a conclusion is given in Section~\ref{s:concl}.

\section{Preliminaries}\label{s:pre}
A constraint satisfaction problem ({\bf CSP}) is defined as a finite set of 
variables ${\cal X} = \{ x_1, \dots , x_n \}$, with domains ${\cal D}= 
\{ D_1, \dots, D_n \}$ associated with them, together with a finite set of 
constraints ${\cal C}$, each on a subset of ${\cal X}$. A CSP $P$ will also be
denoted as $P = ({\cal X}, {\cal D}, {\cal C})$.
A {\bf constraint} $C \in {\cal C}$ is defined as a subset of the Cartesian 
product of the domains of the variables that are in $C$. For instance, 
$C(x_1, x_3,x_4) \subseteq D_1 \times D_3 \times D_4$.
An $n$-uple $(d_1, \dots, d_n) \in D_1 \times \dots \times D_n$ is a
{\bf solution} to a CSP if for every constraint $C \in {\cal C}$ on the 
variables $x_{i_1}, \dots , x_{i_m}$ we have $(d_{i_1}, \dots, d_{i_m}) \in C$.
For finite, linearly ordered domains $D_i$, we define $\min{D_i}$ and 
$\max{D_i}$ to be the minimum value and the maximum value of the domain $D_i$.

We now introduce four notions of local consistency in the order they will be
discussed in the text. Note the use of braces ($\{$, $\}$) and
brackets ($[$, $]$) that indicate a set and an interval of domain values
respectively.

\begin{definition}[Arc consistency]
A binary constraint $C(x_{1},x_2)$ where $D_1$ and $D_2$ are non-empty, is 
called arc consistent iff
$\forall d_1 \in D_1 \; \exists d_2 \in D_2$ such that $(d_1, d_2) \in C$, and
$\forall d_2 \in D_2 \; \exists d_1 \in D_1$ such that $(d_1, d_2) \in C$.
\end{definition}

\begin{definition}[Bound consistency]
An $m$-ary constraint $C(x_{1}, \dots, x_{m})$ where no domain $D_i$
is empty, is called bound consistent iff for each variable $x_i$:
$\forall d_i \in \{\min{D_i}, \max{D_i}\}, \forall j \in \{1, 
\dots ,m \} - \{i\}, \exists d_j \in [\min{D_j}, \max{D_j}]$ such that 
$(d_1, \dots , d_m) \in C$.
\end{definition}

\begin{definition}[Range consistency]
An $m$-ary constraint $C(x_{1}, \dots, x_{m})$ where no domain $D_i$
is empty, is called range consistent iff for each variable $x_i$:
$\forall d_i \in D_i, \forall j \in \{1, \dots ,m \} - \{i\}, 
\exists d_j \in [\min{D_j}, \max{D_j} ]$ such that 
$(d_1, \dots , d_m) \in C$.
\end{definition}

\begin{definition}[Hyper-arc consistency]
An $m$-ary constraint $C(x_{1}, \dots, x_{m})$ where no domain $D_i$
is empty, is called hyper-arc consistent iff for each variable $x_i$:
$\forall d_i \in D_i, \forall j \in \{1, \dots ,m \} - \{i\}, \exists 
d_j \in D_j$ such that $(d_1, \dots , d_m) \in C$.
\end{definition}
In other words, both arc consistency and hyper-arc consistency check whether
any value in every domain does belong to a feasible instance of the constraint,
based on the domains. Range consistency however, does not check the 
feasibility of the constraint with respect to the domains, but with respect 
to intervals that include the domains. It can be regarded as a relaxation of 
hyper-arc consistency. Bound consistency can be regarded as a relaxation of 
range consistency. It does not even check all values in the domains, but only 
the minimum and the maximum value, while still verifying the constraint with 
respect to intervals that include the domains. This is formalized in 
Proposition~\ref{pr:comp}.

\begin{definition}[Consistent CSP]
A CSP is arc consistent if all its binary constraints are. A CSP is range
consistent, respectively, bound consistent or hyper-arc consistent if all its 
constraints are.
\end{definition}

Consider a CSP $P$. If we apply to $P$ an algorithm that achieves range 
consistency  on $P$, we will denote the result as $\Phi_{R}(P)$. Analogously, 
$\Phi_{B}(P)$, $\Phi_{A}(P)$ and $\Phi_{HA}(P)$ denote the achievement of
bound consistency, arc consistency and hyper-arc consistency on $P$ 
respectively.
Let $P_{\emptyset}$ denote a failed CSP, i.e. a CSP with at least one empty
domain. We define a CSP $P=({\cal X}, {\cal D}, {\cal C})$ smaller than a CSP
$P'=({\cal X'}, {\cal D'}, {\cal C'})$ if ${\cal D} \subseteq {\cal D'}$.
This relation is written as $P \preceq P'$. A CSP $P$ is strictly smaller than
a CSP $P'$, i.e. $P \prec P'$, when ${\cal D} \subseteq {\cal D'}$ and 
$D_i \subset D_i'$ for at least one $i$. When both $P \preceq P'$ and
$P' \preceq P$ we write $P \equiv P'$. By convention, $P_{\emptyset}$ is the 
smallest CSP. This notation is adopted from \cite{CDR:comp}.

\begin{proposition}\label{pr:comp}
$\Phi_{HA}(P) \preceq \Phi_{R}(P) \preceq \Phi_{B}(P)$.
\end{proposition}
\begin{proof}
Both hyper-arc consistency and range consistency verify all values of all 
domains. But hyper-arc consistency verifies the constraints with respect to
the exact domains $D_i$, while range consistency verifies the constraints with
respect to intervals that include the domains: 
$[\min{D_i},\max{D_i}]$. A constraint that holds on a domain 
$D_i$ also holds on the interval $[\min{D_i},\max{D_i}]$
since $D_i \subseteq [\min{D_i},\max{D_i}]$. The converse is not
true, see Example~\ref{ex:comp}. Hence $\Phi_{R}(P) \preceq \Phi_{HA}(P)$.

Both range consistency and bound consistency verify the constraints with 
respect to intervals that include the domains. But bound consistency only 
considers $\min{D_i}$ and $\max{D_i}$ for a domain $D_i$, while range
consistency considers all values in $D_i$. Since 
$\{\min{D_i},\max{D_i}\} \subseteq D_i$, 
$\Phi_{B}(P) \preceq \Phi_{R}(P)$. Example~\ref{ex:comp} shows that
$\Phi_{B}(P) \prec \Phi_{R}(P)$ cannot be discarded.
\end{proof}
The following examples clarify Proposition~\ref{pr:comp}.
\begin{example}[Comparing consistencies]\label{ex:comp}
Consider the following CSP:
\begin{displaymath}
P = \left\{
\begin{array}{l}
x_1 \in \{1,3\}, x_2 \in \{2\}, x_3 \in \{1,2,3\}, \\
\verb+alldifferent+(x_1, x_2, x_3).
\end{array} \right.
\end{displaymath}
Then  $\Phi_{B}(P) \equiv P$, while
\begin{displaymath}
\Phi_{R}(P) = \left\{
\begin{array}{l}
x_1 \in \{1,3\}, x_2 \in \{2\}, x_3 \in \{1,3\}, \\
\verb+alldifferent+(x_1, x_2, x_3).
\end{array} \right.
\end{displaymath}
and $\Phi_{HA}(P) \equiv \Phi_{R}(P)$. Next, consider the CSP
\begin{displaymath}
P' = \left\{
\begin{array}{l}
x_1 \in \{1,3\}, x_2 \in \{1,3\}, x_3 \in \{1,3\}, \\
\verb+alldifferent+(x_1, x_2, x_3).
\end{array} \right.
\end{displaymath}
This CSP is obviously inconsistent, since there are only two values available,
namely 1 and 3, for three variables that must be pairwise different.  
$\Phi_{HA}(P')$ will detect this inconsistency, while $\Phi_{R}(P') \equiv P'$.
\end{example}

A useful theorem to derive algorithms that ensure consistency for the \alld\
constraint is Hall's Theorem \cite{hall35}. The following formulation is stated
in terms of the \alld\ constraint. The cardinality of a set $K$ is denoted
by $|K|$.
\begin{theorem}[Hall]\label{thm:hall}
The constraint \verb+alldifferent+$(x_1, \dots, x_n)$ on the variables 
$x_1, \dots, x_n$ with respective domains $D_1, \dots, D_n$ has a solution if 
and only if no subset $K \subseteq \{x_1, \dots ,x_n\}$ exists such that 
$|K| > | \cup_{x_i \in K} D_i |$.
\end{theorem}
As  an application of Theorem~\ref{thm:hall}, let us return to the CSP
$P'$ in Example~\ref{ex:comp}. Take as subset $K = \{x_1, x_2, x_3\}$, then
$|K| = 3$. Furthermore,  $| \cup_{x_i \in K} D_i | = | \{1,3\} | = 2$. For
this subset $K$, Hall's condition does not hold ($3 > 2$), hence this CSP has
no solution.

\section{Local Consistency of a Decomposed CSP}\label{s:dis}
The standard filtering algorithm for the \alld\ constraint is as follows. 
Whenever the domain of a variable contains only one value, remove this value 
from the domains of the other variables that occur in the \alld\ constraint. 
This procedure is repeated as long as possible.
Although this algorithm might seem rather poor or naive, it has been 
successfully implemented in many constraint solvers, for instance in the system 
{\sc Chip} \cite{vanh:chip}.

This filtering algorithm can also be described as follows. A common way to 
rewrite the \alld\ constraint is to generate a sequence of disequalities. For 
instance
\begin{displaymath}
{\tt alldifferent}(x_1, x_2, x_3, x_4) \rightarrow \left\{
  \begin{array}{lll}
  x_1 \neq x_2, &  x_1 \neq x_3, &  x_1 \neq x_4, \\
  x_2 \neq x_3, &  x_2 \neq x_4, &  x_3 \neq x_4. \\
  \end{array}
  \right.
\end{displaymath}
If we apply an algorithm that achieves arc consistency on this set of binary
constraints, we obtain the same filtering as described above. One of the 
drawbacks of this method is the quadratic increase of the number of 
constraints. One needs $\binom{n}{2} = \frac{1}{2}(n^2 - n)$ disequalities to 
express an $n$-ary \alld\ constraint. But an even more important drawback is 
the loss of information. When this set of binary constraints is being made arc
consistent, only two variables are compared at a time. However, when the 
\alld\ constraint is being made hyper-arc consistent, all variables are 
considered at the same time, which gives a much stronger consistency. 
This is shown in Proposition~\ref{pr:AHA}. 
Let $P_{dec}$ denote the decomposed CSP $P$ in which all \alld\ constraints 
have been replaced by a sequence of disequalities.
\begin{proposition}\label{pr:AHA}
$\Phi_{HA}(P) \preceq \Phi_{A}(P_{dec})$.
\end{proposition}
\begin{proof}
Since the definition of arc consistency and hyper-arc consistency is equivalent
for binary constraints, we only need to consider the filtering of the \alld\
constraint. Consider the constraint $C$: $\verb+alldifferent+(x_1,\dots,x_n)$ 
and the corresponding decomposition in terms of disequalities, denoted by 
$C_{dec}$.
If a value $d_i \in D_i$ is not arc consistent w.r.t. the set $C_{dec}$, then 
it is also not hyper-arc consistent w.r.t. $C$. Indeed, when $d_i$ is not arc 
consistent w.r.t. $C_{dec}$, then we cannot find a $d_j \in D_j$ for some 
variable $x_j$ such that $x_i \neq x_j$. But then we also cannot find an
$n$-uple $(d_1, \dots, x_n) \in C$, since we cannot find a value $d_j \in 
D_j$ such that $d_i \neq d_j$. Therefore, $\Phi_{HA}(P) \preceq 
\Phi_{A}(P_{dec})$. The converse is not true, as illustrated in 
Example~\ref{ex:AHA}.
\end{proof}
\begin{example}[Hyper-arc and arc consistency compared]\label{ex:AHA}
For some integer $n \geq 3$, consider the CSP's
\begin{displaymath}
P = \left\{ 
\begin{array}{l}
x_1 \in \{1,\dots, n-1\}, \dots,  x_{n-1} \in \{1,\dots, n-1\}, 
x_n \in \{1,\dots, n \}, \\
\verb+alldifferent+(x_1, \dots, x_n)
\end{array} \right.
\end{displaymath}
\begin{displaymath}
P_{dec} = \left\{ 
\begin{array}{l}
x_1 \in \{1,\dots, n-1\}, \dots,  x_{n-1} \in \{1,\dots, n-1\}, 
x_n \in \{1,\dots, n \}, \\
x_1 \neq x_2, \dots, x_{n-1} \neq x_n.
\end{array} \right.
\end{displaymath}
Now $\Phi_{A}(P_{dec}) \equiv P_{dec}$, while
\begin{displaymath}
\Phi_{HA}(P) = \left\{
\begin{array}{l}
x_1 \in \{1,\dots, n-1\}, \dots,  x_{n-1} \in \{1,\dots, n-1\}, 
x_n \in \{ n \}, \\
\verb+alldifferent+(x_1, \dots, x_n).
\end{array} \right.
\end{displaymath}
\end{example}

Our next goal is to find a consistency notion for the set of disequalities 
that is equivalent to the hyper-arc consistency notion for the \alld\ 
constraint. Relational consistency can be used for this.

\begin{definition}[Relational $(1,m)$ consistency, \cite{dechter97}]
A set of constraints $S = \{C_1, \dots, C_m\}$ is {\em relationally
$(1,m)$-consistent} iff all domain values $d \in D_i$ of variables appearing
in $S$, appear in a solution to the $m$ constraints, evaluated simultaneously.
A CSP $P = ({\cal X, D, C})$ is relationally $(1,m)$-consistent iff every set 
of $m$ constraints $S \subseteq {\cal C}$ is relationally $(1,m)$-consistent.
\end{definition}
Note that arc consistency is equivalent to $(1,1)$-consistency.

Again, let $P$ be the CSP that consists only of the \alld\ constraint and a
corresponding set of variables and domains.
\begin{proposition}
$\Phi_{HA}(P) \equiv \Phi_{R(1, \frac{1}{2}(n^2 - n))C}(P_{dec})$.
\end{proposition}
\begin{proof}
By construction we have that the \alld\ constraint is equivalent to the 
simultaneous consideration of the sequence of corresponding disequalities.
The number of disequalities is precisely $\frac{1}{2}(n^2 - n)$. If we consider
only $\frac{1}{2}(n^2 - n) - i$ disequalities simultaneously 
($1 \leq i \leq \frac{1}{2}(n^2 - n) - 1$), there are $i$ unconstrained 
relations between variables, and the corresponding variables could take the
same value when a certain instantiation is considered. Therefore, we really 
need to take all $\frac{1}{2}(n^2 - n)$ constraints into consideration, which 
corresponds to the relational $(1, \frac{1}{2}(n^2 - n))$-consistency.
\end{proof}

As suggested before, the pruning performance of $\Phi_{A}(P_{dec})$ is rather
poor. Moreover, the complexity is relatively high, namely around $O(n^2)$, 
whereas the hyper-arc consistency algorithms are around $O(dn^{1.5})$, where 
$d$ is the maximum cardinality of the domains and $n$ is the number of 
variables involved \cite{leconte96,regin}.
Nevertheless, this filtering algorithm applies quite well to several 
problems, such as the $n$-queens problem  ($n<200$) \cite{leconte96,puget98}.

Other work on the comparison of the \alld\ constraints and the corresponding
decomposition has for instance been done in \cite{stergiou99} and 
\cite{gent2000}.

\section{Bound Consistency}\label{s:bound}
The notion of bound consistency for the \alld\ constraint was introduced
by Puget \cite{puget98}. We summarize his method in this section. Puget
uses Hall's Theorem to construct an algorithm that achieves bound consistency.

\begin{definition}[Hall interval]
Given an interval $I$, let $K_I$ be the set of variables $x_i$ such
that $D_i \subseteq I$. We say that $I$ is a Hall interval iff 
$|I| = |K_I|$.
\end{definition}

\begin{proposition}[Puget \cite{puget98}]\label{pr:bound}
The constraint \verb+alldifferent+$(x_1, \dots, x_n)$ where no domain $D_i$ is
empty, is bound consistent iff 
\begin{itemize}
\item for each interval $I$: $|K_I| \leq |I|$, 
\item for each Hall interval $I$: $\{ \min{D_i}, \max{D_i} \} \cap I = 
\emptyset$ for all $x_i \notin K_I$.
\end{itemize}
\end{proposition}

Proposition~\ref{pr:bound} can be used to construct an algorithm that achieves
bound consistency on the \alld\ constraint. Indeed, we could check every 
interval $I$ with bounds ranging from the minimum of all domains to the 
maximum of all domains. 
When $|I| \leq |K_I|$, we know that the constraint is
inconsistent. And for each Hall interval, we remove all $\min{D_i}$ and
$\max{D_i}$ until $\{ \min{D_i}, \max{D_i} \} \cap I 
= \emptyset$. Puget gives an implementation with the time complexity 
of $O(n\log{n})$.

In \cite{mehlhorn}, Mehlhorn and Thiel present an algorithm that achieves bound
consistency of the \alld\ constraint in time $O(n)$ plus the time required for
sorting the interval endpoints. In particular, if the endpoints are from a 
range of size $O(n^k)$  for some constant $k$, the algorithm runs in linear
time.

\begin{example}\label{ex:bound}
The following simple problem shows an application of the bound consistency 
algorithm based on intervals. 
\begin{displaymath}
P = \left\{ 
\begin{array}{l}
x_1 \in \{1,2\}, x_2 \in \{1,2\}, x_3 \in \{2,3\}, \\
\verb+alldifferent+(x_1, x_2, x_3).
\end{array} \right.
\end{displaymath}
Intuitively, observe that the variables $x_1$ and
$x_2$ both have domain $\{ 1, 2 \}$. So these two variables together range
over two values, and for a feasible instantiation they must be different. This
means that the values 1 and 2 must be assigned to these two variables. Hence,
values 1 and 2 cannot be assigned to any other variable and therefore, 
value 2 will be removed from the domain of $x_3$.

The algorithm detects this when the interval $I$ is set to $I = \{1,2\}$. Then 
the number of variables for which $D_i \subseteq I$ is 2, namely $x_1$ and
$x_2$. Since $|I|=2$, $I$ is a Hall interval. The domain of $x_3$ is not in
this interval, and $\{ \min{D_3}, \max{D_3} \} \cap I = \{ \min{D_3} \} $. 
In order to obtain the empty set in the right hand side of the last equation, 
we need to remove $\min{D_i}$. The resulting CSP is bound consistent.
\end{example}

\section{Range Consistency}\label{s:range}
An algorithm that achieves range consistency was introduced by Leconte
\cite{leconte96}. We follow the same procedure as in the previous example.
Leconte also uses Hall's Theorem to construct the algorithm. 

\begin{definition}[Hall set]
Given a set of variables $K$, let $I_K$ be the interval $[\min{D_K},
\max{D_K}]$, where $D_K = \cup_{x_i \in K} D_i$. We say that $K$ is a Hall 
set iff $|K| = |I_K|$. 
\end{definition}
Note that in the above definition $I_K$ does not necessarily need to be a Hall
interval.

\begin{proposition}[Leconte \cite{leconte96}]\label{pr:range}
The constraint \verb+alldifferent+$(x_1, \dots, x_n)$ where no domain $D_i$ is
empty, is range consistent iff for each Hall set $K \subseteq \{ x_1, \dots 
x_n \}$: $D_i \cap I_K = \emptyset$ for all $x_i \notin K$.
\end{proposition}

We can deduce an algorithm from Proposition~\ref{pr:range} in a similar way as
we did for the algorithm for bound consistency. Leconte implemented an 
algorithm that achieves range consistency with a complexity of $O(n^2 d)$, 
where $d$ is the average size of the domains.

Observe that this algorithm is similar to the algorithm for bound consistency.
Where the algorithm for bound consistency takes the domains as a starting 
point, the algorithm for range consistency takes the variables. But they both 
attempt to reach a situation in which the cardinality of a set of variables is
equal to the cardinality of the union of the corresponding domains, as was 
illustrated in Example~\ref{ex:bound}.

\section{Hyper-arc Consistency}\label{s:arc}
A filtering algorithm that achieves hyper-arc consistency for constraints of
difference was proposed by R\'egin \cite{regin}. A similar result was
obtained independently by Costa \cite{costa}. Before we can 
introduce this algorithm, we have to establish a connection with the maximum 
matching problem in graph theory. The standard reference to matching theory is
the book by Lov\'asz and Plummer~\cite{LP:matching}.

\subsection{Connections with Matching Theory}
Consider again the scheduling problem from Example~\ref{ex:speech}. To
illustrate the problem, assume that Krzysztof and Luca decided not to speak.
We now want to model this problem graph-theoretically. First we introduce the
definition of a bipartite graph.

\begin{figure}
\centerline{\psfig{figure=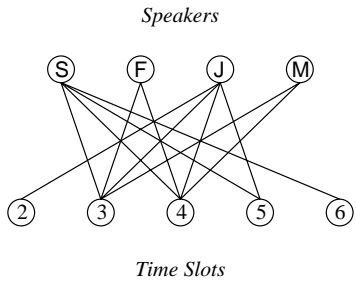}}
\caption{The value graph for the revised speech scheduling problem}
\label{fig:graph}
\end{figure}

\begin{definition}[Bipartite graph]
A {\em graph} $G$ consists of a finite non-empty set of elements $V$ called 
{\em nodes} and a set of pairs of nodes $E$ called {\em edges}. If the node 
set $V$ can be partitioned into two disjoint non-empty
sets $X$ and $Y$ such that all edges in $E$ join a node from $X$ to a node
in $Y$, we call $G$ {\em bipartite} with {\em bipartition} $(X,Y)$. We also 
write $G=(X,Y,E)$.
\end{definition}
The remaining speakers from Example~\ref{ex:speech} and their available times 
can be represented by the bipartite graph in Figure~\ref{fig:graph}.
Both speakers and time periods are represented by nodes, and these two sets
of nodes are connected by edges, giving the bipartition 
$({\rm\it Speakers},{\rm\it Times})$.
We call the constructed bipartite graph of an \alld\ constraint $C$ the 
{\em value graph} of $C$. Let $X_C$ denote the variables occurring in a 
constraint $C$, with corresponding domains $D_C$.

\begin{definition}[Value graph]
Given an \verb+alldifferent+ constraint $C$, the bipartite graph 
$GV(C)=(X_C, D_C, E)$ where $(x_i, d) \in E$ iff $d \in D_i$ is called 
the value graph of $C$.
\end{definition}
\begin{definition}[Maximum matching]
A subset of edges in a graph $G$ is called a {\em matching} if no two edges 
have a node in common. A matching of maximum cardinality is called a 
{\em maximum matching}. A matching $M$ {\em covers a set} $X$ if every node
in $X$ is an endpoint of an edge in $M$.
\end{definition}
Note that a matching that covers the set of speakers in Figure~\ref{fig:graph}
is a maximum matching. The following theorem gives the link between a maximum 
matching in a bipartite graph and hyper-arc consistency of the \alld\ 
constraint.
\begin{proposition}[R\'egin \cite{regin}]\label{pr:hyper}
The constraint $C:$ \verb+alldifferent+$(x_1, \dots, x_n)$ is hyper-arc
consistent iff every edge in its value graph $GV(C)$ belongs to a matching 
which covers $X_C$ in $GV(C)$.
\end{proposition}
\begin{figure}
\centerline{\psfig{figure=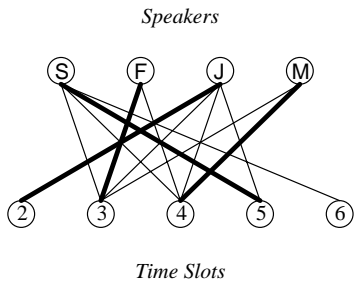}}
\caption{A maximum matching in the value graph}
\label{fig:graph_match}
\end{figure}
\begin{figure}
\centerline{\psfig{figure=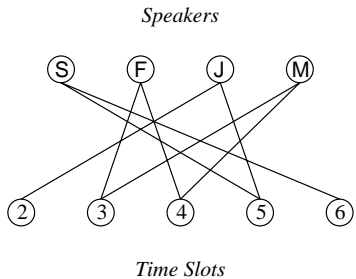}}
\caption{The value graph after filtering}
\label{fig:graph_filt}
\end{figure}
An illustration of Proposition~\ref{pr:hyper} is given in 
Figures~\ref{fig:graph_match} and~\ref{fig:graph_filt}.
The fat lines in the graph of Figure~\ref{fig:graph_match} denote a maximum 
matching that covers all speaker nodes. Not all edges belong to such a 
matching, and by Proposition~\ref{pr:hyper} they can be removed. When these 
edges are removed, the resulting \alld\ constraint is hyper-arc consistent.
This is depicted in Figure~\ref{fig:graph_filt}, which corresponds to
Table~\ref{tb:times_HA}.
\begin{table}
\begin{center}
\caption{Filtered time slots for the speakers}
\label{tb:times_HA}
\begin{tabular}{|l|c|} \hline \hline 
Speaker      & Available \\ \hline
Sebastian    & $\{5,6\}$  \\ 
Fr\'ed\'eric & $\{3,4\}$  \\ 
Jan-Georg    & $\{2,5\}$  \\ 
Maarten      & $\{3,4\}$  \\ \hline \hline
\end{tabular}
\end{center}
\end{table}

\subsection{An Algorithm for Achieving Hyper-arc Consistency}
An algorithm that achieves hyper-arc consistency for the \alld\ constraint 
should remove all those edges in the corresponding value graph that do not 
belong to a maximum matching. Berge has given a property that identifies
exactly these edges \cite{berge73}. But first, we introduce some definitions
we need for this property.
\begin{definition}
Let $M$ be a matching in a graph $G = (V, E)$.
An {\em alternating path} or {\em alternating cycle} is a path or a cycle 
whose edges are alternately in $M$ and in $E-M$. The {\em length} of a path or
a cycle is the number of edges it contains. A node is called {\em free} w.r.t.
$M$ if it is not incident to a matching edge.
\end{definition}
For instance, in Figure~\ref{fig:graph_match}, $(3,F,4,M,3)$ is an even 
alternating cycle of length 4. Node 6 is a free node. 
\begin{proposition}[Berge]
An edge belongs to a maximum matching iff for some maximum matching, it 
belongs to either an even alternating path which begins at a free node, or
to an even alternating cycle.
\end{proposition}
With this property, we are able to identify and remove edges that are not in
any maximum matching. Note that we need to construct a maximum matching 
before we can apply this property. The algorithm that achieves hyper-arc 
consistency is represented in Figure~\ref{fig:hyper}.
\begin{figure}
\footnotesize
\begin{tabbing}
MMMM\=MM\= \kill
\> Input: constraint of difference $C$, variables ${\cal X}$ and domains
${\cal D}$ \\
\> Output: false when no solution, otherwise true and updated domains \\
\> begin \\
\> 1 \> Build $GV = (X_C, D_C, E)$\\
\> 2 \> $M(GV) \leftarrow$ {\sc ComputeMaximumMatching}$(GV)$\\
\> 3 \> if $|M(GV)| < |X_C|$ then return false \\
\> 4 \> {\sc RemoveEdgesFromG}$(GV,M(GV))$ \\
\> 5 \> return true \\
\> end
\end{tabbing}
\normalsize
\caption{An algorithm for achieving hyper-arc consistency}
\label{fig:hyper}
\end{figure}
To construct the value graph $GV$, we need $O(d|X_C| + |X_C| + |D_C|)$ steps,
where $d$ is the maximum cardinality of a variable domain.
The procedure {\sc ComputeMaximumMatching}$(GV)$ computes a maximum matching in
the graph $GV$. This can be done for instance with a so-called 
{\em augmenting path} algorithm. Hopcroft and Karp gave an 
implementation for this that runs in $O(\sqrt{|X_C|}m)$ time, where $m$ is the 
number of edges of $GV$ \cite{HK:maximum}. Their
algorithm still remains essentially the best known \cite{Cookea:CO}.

From Hall's Theorem we already know that whenever we find a subset of nodes 
the cardinality of which exceeds the cardinality of the corresponding set of 
domain values, no matching exists that saturates $X_C$.
This is checked in line 3. In the procedure {\sc RemoveEdgesFromG}$(GV,M(GV))$
the actual filtering takes place. Instead of applying Berge's property 
directly, we can translate the problem in such a way, that we have to search 
for the so-called strongly connected components of the graph \cite{regin}. For
this problem we can use an implementation by Tarjan that runs in $O(n+m)$ time
on graphs with $n$ nodes and $m$ edges \cite{regin,tarjan72}. In the algorithm
from Figure~\ref{fig:hyper}, the search for a maximum matching remains the 
dominant factor, hence the total algorithm runs in $O(\sqrt{|X_C|}m)$ time.

The notion of hyper-arc consistency was introduced by Mohr and Masini 
\cite{MM:good}. They also give a general algorithm to achieve this notion. For
an $n$-ary \alld\ constraint, where the domain size of all variables is bounded
by $d$, $D_i \leq d$, the time complexity of the general algorithm is
$O(\frac{d!}{(d-n)!})$, whereas the time complexity of the above algorithm is
$O(dn\sqrt{n})$.

\section{Conclusions and Future Work}\label{s:concl}
In this paper, an overview of several filtering techniques for the \alld\ 
constraint has been given. A comparison of these different techniques has been 
made by means of corresponding notions of local consistency and algorithms to 
achieve them.

However, there are other interesting articles related to this subject, that
are not considered in this paper. For instance, Focacci et al. \cite{focacci99}
use information from the \alld\ constraint for a filtering technique based on 
reduced costs. Furthermore, in \cite{regin99} R\'egin introduced the symmetric
\alld\ constraint, together with filtering algorithms for this constraint. \
Finally, Bart\'ak considers a dynamic version of the \alld\ constraint 
\cite{bartak01}.
\bibliography{biblhoeve}

\end{document}